\documentclass{article}
\usepackage[T1]{fontenc} 
\usepackage[utf8]{inputenc} 
\usepackage{ismir,amsmath,cite,url}
\usepackage{graphicx}
\usepackage{color}

\usepackage{booktabs} 
\usepackage[bookmarks=false,hidelinks]{hyperref}
\usepackage{tabularx}
\usepackage{float}
\usepackage{xcolor}
\usepackage{multicol}
\usepackage{multirow}
\usepackage{stfloats} 
\usepackage{caption}
\usepackage{siunitx}
\usepackage{doi}

\newcommand{\ch}[1]{{\color{black} #1}}
\newcommand{\rev}[1]{{\color{black} #1}} 

\usepackage{xspace}
\newcommand{\repo}{\url{https://github.com/CPJKU/mpteval}}

\usepackage{lineno}

\title{Towards Musically Informed Evaluation\\of Piano Transcription Models}

\multauthor
{Patricia Hu$^1$ \hspace{0.54cm} Luk\'a\v{s} Samuel Mart\'ak$^{1,2}$ \hspace{0.54cm} Carlos Cancino-Chac\'on$^1$ \hspace{0.54cm} Gerhard Widmer$^{1,2}$} 
{
$^1$ Institute of Computational Perception, Johannes Kepler University Linz, Austria\\
$^2$ {LIT AI Lab, Linz Institute of Technology, Austria}\\
{\tt patricia.hu@jku.at}
}

\def\authorname{P. Hu, L. Mart\'ak, C. Cancino-Chac\'on and G. Widmer}

\begin{document}

\maketitle
\begin{abstract} 
Automatic piano transcription models are typically evaluated using simple frame- or note-wise information retrieval (IR) metrics. Such benchmark metrics do not provide insights into the transcription quality of specific musical aspects such as articulation, dynamics, or rhythmic precision of the output, which are essential in the context of expressive performance analysis.
Furthermore, in recent years, MAESTRO has become the de-facto training and evaluation dataset for such models. However, inference performance has been observed to deteriorate substantially when applied on out-of-distribution data, thereby questioning the suitability and reliability of transcribed outputs from such models for specific MIR tasks.
In this work, we investigate the performance of three state-of-the-art piano transcription models in two experiments. In the first one, we propose a variety of musically informed evaluation metrics which, in contrast to the IR metrics, offer more detailed insight into the musical quality of the transcriptions.
In the second experiment, we compare inference performance on real-world and perturbed audio recordings, and highlight musical dimensions which our metrics can help explain.
Our experimental results highlight the weaknesses of existing piano transcription metrics and contribute to a more musically sound error analysis of transcription outputs.
\end{abstract}


\section{Introduction}\label{sec:introduction}


Automatic Music Transcription (AMT) refers to the task of converting audio signals into symbolic music representations. The target output format can be a full symbolic score including quantized rhythm, time signature and pitch spelling information, or a mid-level physical MIDI(-like) representation, describing notes in terms of their onset and offset times, pitch and velocity \cite{hawthorne2017onsets, kong2021highresolution, hawthorne2021sequence}.

AMT methods are typically evaluated using information retrieval (IR) metrics like precision, recall and F1 score \cite{bay2009evaluation}. These IR metrics can be computed at the level of frames, by comparing binary piano roll-like matrices, or at the level of note lists, by comparing notes in terms of their onset, offset, pitch and/or velocity attributes. Each error (i.e. misplaced frame or note activity) has equal weight, resulting in limited explanatory power of these metrics with respect to the underlying musical material \cite{hawthorne2017onsets, ycart2020investigating}.

As in many other areas in MIR, the current state of the art is defined by deep neural networks \ch{\cite{lu2023multitrack, hawthorne2021sequence, kong2021highresolution, maman2022unaligned}}. To a large extent, this progress has been enabled by the release of the MAESTRO dataset \cite{hawthorne2019enabling}, which made well-aligned audio-MIDI piano performance data available
on a large scale. The most up-to-date version of the MAESTRO dataset\footnote{https://magenta.tensorflow.org/datasets/maestro} contains close to 200 hours of performance data from close to 1300 recordings of Western classical piano repertoire. State-of-the-art piano transcription systems achieve beyond 90\% frame-level, or 80\% note-level F1 scores on its test split \cite{hawthorne2019enabling, kong2021highresolution, hawthorne2021sequence, gardner2021mt3}, and have led to the release of large-scale transcribed solo piano performance datasets \cite{kong2020giantmidi, zhang2023atepp}.



Although these results are impressive, we believe that two important aspects have been largely overlooked: first, the validity and (lack of) explanatory power of the standard evaluation metrics with respect to musically relevant information, and second, the reliability of these transcription models on out-of-distribution data. In this work, we address the first problem by proposing a set of musically informed evaluation metrics that support a more nuanced understanding of piano transcription errors. The metrics are intended to be used in the context of computational performance studies,
and therefore focus on musical dimensions that are commonly studied in the context of expressive piano performance analysis and generation. We demonstrate our metrics on a subset of the MAESTRO dataset, which we transcribe using three state-of-the-art transcription models. In particular, we contrast the performance of these models, as evaluated with the standard IR metrics, with their performance on musical dimensions such as timing, articulation and dynamics which we can evaluate using our set of musically informed metrics. 

Then, to elucidate the second problem, we re-record a subset of the MAESTRO dataset on a Yamaha Disklavier grand piano and further manipulate the audio recordings by adding different levels of noise and reverberation. 
An analysis of the outputs of these trained transcription models on these recordings provides some detailed insights into the lack of generalization on out-of-distribution data. We make our set of metrics, data and all experimental results available at \repo.


\section{Related Work}\label{sec:related_work}

This section briefly reviews the standard IR evaluation metrics along with criticism related to these, followed by a description of the benchmark datasets typically used for evaluating transcription methods.

Precision, recall and F1 score are the standard evaluation metrics used in AMT \cite{hawthorne2017onsets, kong2021highresolution, hawthorne2021sequence, bay2009evaluation}. They can be computed either at the level of frames or at the level of notes. \ch{For frame-level evaluation, two binary piano roll matrices $M, \hat{M} \in \{0, 1\}^{P \times T}$ are compared, where $p=1,...,P$ defines the pitch range and $t=1,...,T$ the time step (typically with a resolution of 10ms \cite{bay2009evaluation}). Both $M$ and $\hat{M}$ are sparse matrices where 1 at a given index $p,t$ indicates that a note with pitch $p$ is active at time frame $t$.}

Note-level metrics are computed by comparing lists of notes, in which each note is described by a tuple describing the onset, offset, pitch, and (where predicted) velocity. Note-based metrics can be based on onset information only, onset and offset information (i.e., note durations), or on predicted onset, offset and velocity. In onset-only note evaluation, a note is considered correct if its onset falls within a $\pm$50 ms threshold of its respective target onset. For onset-and-offset note-level evaluation, the note offset must fall within the greater of either an offset tolerance threshold of $\pm$50 ms, or a duration threshold 20\% of the ground truth duration \cite{bay2009evaluation}. If velocity is included in the evaluation, an estimated note is considered correct if its velocity (after some normalization and rescaling operations) falls within a 0.1 tolerance threshold of the velocity of the corresponding reference note \cite{hawthorne2017onsets}.

The need for better (i.e., musically or perceptually sound) transcription metrics has been expressed by various researchers before. Hawthorne et al.~\cite{hawthorne2017onsets} point out that frame- and onset-only note-level evaluation does not sufficiently capture musically relevant information. Similarly, Ycart et al. \cite{ycart2020investigating} and Daniel et al. \cite{daniel2008perceptually} focus on the problem of perceptual saliency of different kinds of transcription errors and each propose a new, perceptually (more) valid transcription metric. Finally, McLeod and Steedman \cite{mcleod2018evaluating} focus on the problem of audio-to-score transcription and propose a new metric that jointly evaluates voice separation, metrical alignment, note value detection and harmonic analysis along with multi-pitch detection.




With respect to training and evaluation data for solo piano transcription, until the introduction of MAESTRO \cite{hawthorne2019enabling}, MAPS \cite{emiya2010maps} was used as the standard dataset. Apart from size, the biggest difference between the two is the diversity of the captured recording environments: while MAESTRO exclusively contains Disklavier recordings from the Yamaha International Piano e-Competition\footnote{http://piano-e-competition.com/}, MAPS contains Disklavier recordings and synthesized audio simulating various recording environments. 
The prevailing trend in evaluating current piano transcription models centers around the MAESTRO dataset \cite{kong2021highresolution, hawthorne2021sequence, lu2023multitrack}, and most models that do include MAPS in their evaluation \cite{hawthorne2017onsets, hawthorne2021sequence} use the split proposed in \cite{sigtia2016end}, which only includes Disklavier recordings in the test split. 
Both frame- and note-level metrics are usually computed for each piece in a given test set, and their mean is subsequently reported as the inference performance for a given model and dataset/split. Frame-level metrics are typically higher than note-level ones due to common known transcription errors such as merged or segmented notes.

\section{Musically informed metrics}\label{sec:musical_metrics}

In this section we describe our proposed metrics that are meant to capture different musical dimensions commonly studied in the context of expressive performance. 
Each metric compares a ground truth to a predicted MIDI performance by measuring the Pearson correlation between a performance parameter computed from the ground truth and from the predicted MIDI, respectively. We choose a correlational measure to ensure all metrics fall into the same range. The goal is to quantify dimensions of musical quality of transcriptions that are otherwise obscured by standard IR metrics. In particular, we wish to capture dimensions that are important for computational performance studies that make use of automatically transcribed piano performances.

\subsection{Timing}
Timing can be described as expressive deviations from the metrical grid. 
A common measure of expressive timing in computational performance analysis is the inter-onset-interval (IOI), that is, the amount of time passed between two consecutive notes belonging to the same stream.\footnote{We use the term \emph{stream} as a generalization of the concept of a voice in polyphonic music~\cite{temperley:2009jnmr}.} 

To evaluate how well a transcription preserves the micro onset deviations, we predict a monophonic melody line and the accompaniment part (i.e., all notes not belonging to the melody line) in a given MIDI using the skyline algorithm for melody identification \cite{uitdenbogerd1999melodic}.\footnote{\ch{The skyline algorithm has been shown to be very competitive in identifying melody lines in Western classical piano music, even when compared to more recent machine learning-based algorithms.
(e.g., see Figure 5 in \cite{simonetta2019convolutional}).}}

Then we compute the IOIs of these streams both on the ground truth and predicted performance, and measure their correlation.
Note that for non-strictly monophonic streams (like the accompaniment part), the IOI between notes that belong to the same onset (i.e., chords) is zero.
The result of this process gives us two measures, which we call \emph{Melody IOI} and \emph{Accompaniment IOI}.

\subsection{Articulation}
Articulation in expressive piano performance refers to how (adjacent) notes are played in terms of their duration, intensity, and clarity, resulting in expressive strategies such as \textit{legato}, \textit{staccato} or \textit{marcato}. 
Computationally, articulation is measured as the ratio between the time interval from the offset of the current note to the onset of the next note, and the time between the onsets of the two notes.
\cite{bresin2000articulation, repp1995legato, drake1993accent}.

We use the skyline algorithm \cite{uitdenbogerd1999melodic} to extract monophonic melody and bass lines within a performance, and compute a sequence of KOR values, for each pair of successive note events, for both the target and the predicted performance MIDI for both streams, and their ratio.
We define three metrics for capturing articulation:
\begin{enumerate}
\item \emph{Melody KOR}: the correlation between the KOR sequences of the melody lines of the ground truth and the predicted performance MIDI.
\item \emph{Bass KOR}: the correlation between the KOR sequences of the bass lines of the ground truth and the predicted performance MIDI.
\item \emph{Ratio KOR}: for this metric, we consider the ratio of KOR sequences of the melody to the bass line. A ratio KOR greater than 1 indicates that the melody voice is played more legato than the bass voice.
The Ratio KOR metric is computed as correlation of this ratio between the ground truth and the predicted performance MIDI. 
\end{enumerate}

\subsection{Harmony}
Aspects such as harmonic tension have been shown to be determining factors for various performance decisions (particularly relating to expressive tempo and dynamics~\cite{cancino2018computational, herremans2019towards}). 
To quantify how well harmonic tension is preserved in a transcription, we use two features proposed by Herremans and Chew~\cite{herremans2016tension} based on Chew's spiral array model \cite{chew2016playing}. 
This model is a three dimensional representation of pitch classes, chords and keys constructed in such a way that spatial proximity represents close tonal relationships.\footnote{\ch{We chose this model for its simplicity and music-theoretical grounding. Note that these features were designed for Western tonal music and may be less effective in capturing tension in other types of music.}}
We use two metrics to capture the preservation of harmonic tension:

\ch{
\begin{enumerate}
\item \emph{Cloud Diameter}: this metric measures the maximal tonal distance as the maximum dispersion between notes in a musical segment
\item \emph{Cloud Momentum}: this metric captures the harmonic movement in a segment as the tonal distance between consecutive sections.
\end{enumerate}

For both metrics, we compute the respective feature on overlapping windows 
for both the ground truth and transcribed MIDI, and measure their correlation.
} 

\subsection{Dynamics}
For comparing the performance of transcription models regarding expressive dynamics, we use the loudness ratio of the melody and bass lines as a proxy to identify how well a transcription preserves the dynamics of the performance. 
We estimate the loudness as the ``energy'' of a stream (i.e., melody or bass line), which is computed using the MIDI velocity following a model proposed by Dannenberg~\cite{dannenberg2006interpretation}. 
The loudness ratio is then computed as follows (cf. Equation 8 in \cite{dannenberg2006interpretation}):
\begin{equation}
\text{R}(t) = \log\left( \frac{m \cdot \text{vel}_{mel}(t) + b}{m \cdot \text{vel}_{bass}(t) + b} \right)
\end{equation}

\noindent where $\text{vel}_{mel}(t)$ and $\text{vel}_{bass}(t)$ are the MIDI velocities of the melody and bass lines at time $t$, respectively, and $m$ and $b$ are constant parameters that depend on the dynamic range of the audio signal.
We compute the loudness ratio for both the ground truth and estimated performance MIDI, and compute the correlation between these ratios as our metric for dynamics.\footnote{\ch{Dannenberg's model was chosen for its simplicity, relying only on MIDI velocity and dynamic range (note that parameters $m$ and $b$ cancel each other out when computing the correlation of the loudness ratio).}}



\section{Data}\label{sec:data}

\begin{table}[t]
    \small
    \centering
    \begin{tabular}{lrrr}
\toprule
\textbf{composer} & \textbf{pieces} & \textbf{performances} & \textbf{duration (min)} \\
\midrule
Bach & 1 & 7 & 23.36 \\
Beethoven & 5 & 28 & 285.54 \\
Chopin & 4 & 15 & 150.28 \\
Debussy & 2 & 3 & 32.06 \\
Glinka & 1 & 2 & 10.35 \\
Liszt & 3 & 12 & 58.98 \\
Mozart & 1 & 2 & 29.02 \\
Rachmaninoff & 2 & 3 & 11.87 \\
Haydn & 3 & 9 & 90.23 \\
Schubert & 3 & 17 & 107.27 \\
Scriabin & 1 & 5 & 55.05 \\
\midrule
\textbf{Total} & \textbf{26} & \textbf{103} & \textbf{854.01} \\
\bottomrule
\end{tabular}

    \captionsetup{width=.5\textwidth}
    \caption{\ch{Overview of chosen composers, pieces, and performances in the MAESTRO subset in our evaluation set.}}
    \label{tab:eval_dataset_table}
\end{table}

\ch{
For our experiments, we create an evaluation set with three subsets:

\begin{enumerate}
    \item \textit{MAESTRO}: We select audio recordings from the MAESTRO dataset, covering a diverse range of musical repertoire, composers, and performers, using all (train, validation, and test) splits as provided by the authors \cite{hawthorne2021sequence}. This choice tests whether the split category affects model generalization. \footnote{The official MAESTRO splits \cite{hawthorne2019enabling} ensure a unique piece-to-split mapping.} An overview of the selected subset is shown in Table \ref{tab:eval_dataset_table}.
    \item \textit{Disklavier}: We re-record 
    our MAESTRO subset on a Yamaha Disklavier Enspire ST C1X
    using the Focusrite Scarlett 18i8 and a pair of AKG P420 microphones in a moderately bright, fully carpeted room with asymmetric geometry and low background noise level.
    \item \textit{revnoise}: To simulate more challenging real-world environments, we further add perturbations using different levels of reverberation and noise (see Section \ref{sec:ood_inference}) on selected recordings from both the \textit{MAESTRO} and \textit{Disklavier} subsets.
\end{enumerate}

}

We compare three state-of-the-art piano transcription models: Onsets and Frames \cite{hawthorne2019enabling} and the Transformer transcription model \cite{hawthorne2021sequence} by Google/Magenta (which we will refer to as OaF and T5 respectively, in the following), and the high-resolution onset and offset regression model by Bytedance \cite{kong2021highresolution} (referred to as Kong).

We transcribe all the recordings in our MAESTRO, \textit{Disklavier}, and \textit{revnoise} subsets using the (officially provided) trained models, and these transcriptions then form our evaluation set. 
Note that all audio recordings from the \textit{Disklavier} and \textit{revnoise} subsets are only used for testing; we use the MAESTRO-trained models as they are provided by the respective authors via their repositories.



\begin{figure*}[b]
  \includegraphics[alt={Model performance comparison as evaluated on note-offset F1 score and our proposed musical metrics, by composer.},width=\textwidth]{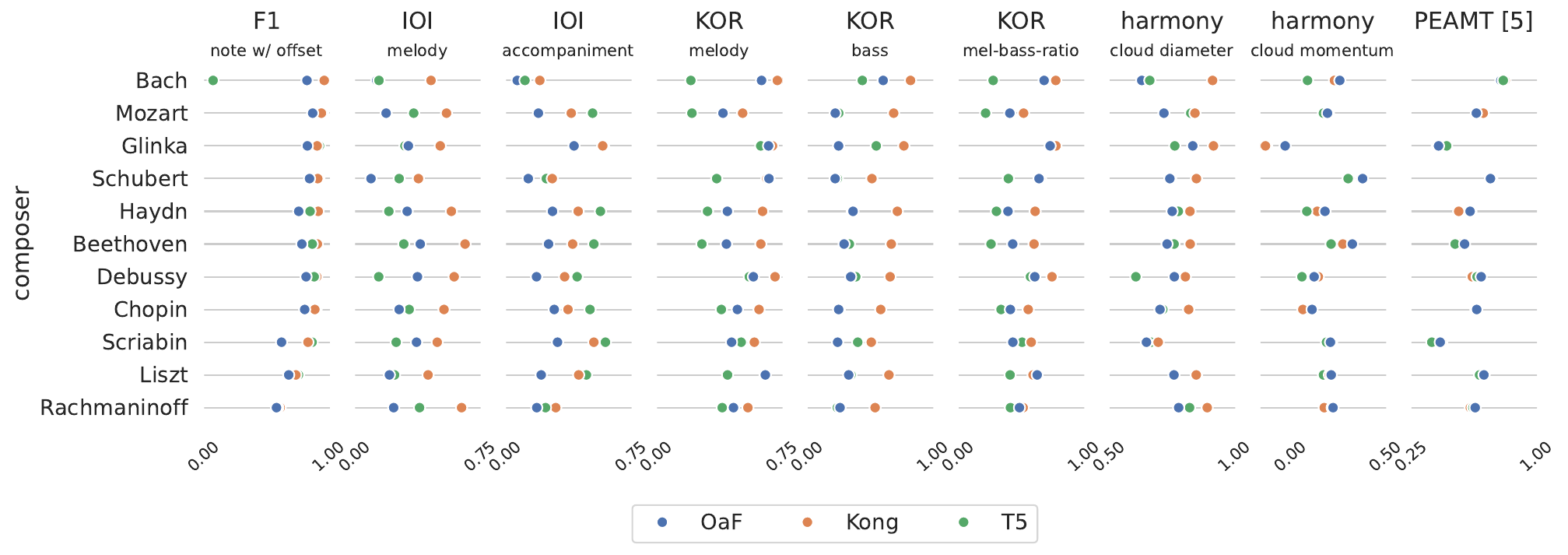}
  \caption{Model performance comparison as evaluated on note-offset F1 score and our proposed musical metrics, by composer.}
  \label{fig:composerwise_MAESTRO}
\end{figure*}

\section{Demonstration of musically informed metrics}\label{sec:demonstration}

We now discuss the experimental results obtained with the three systems and explain the relation between our metrics and the standard IR metrics as computed on transcriptions of the MAESTRO subset of our evaluation set. We focus in particular on the musical dimensions that can be better understood through our metrics. For the standard evaluation, we include the frame-level score, and all note-level F1 scores other than the onset-only one as it does not capture offset and velocity information. We compute the note-level metrics with the official \texttt{mir\_eval} python implementation \cite{raffel2014mir_eval}.

We start our discussion with Table \ref{tab:maestro_all_metrics_res}, which summarizes the evaluation results per model and metric.
For comparative reasons, we also include a perceptually informed piano transcription metric, PEAMT \cite{ycart2020investigating}.

\begin{table}[ht]
    \centering
    \begin{tabular}{lrrr}
\toprule
\textbf{Metric} & \textbf{OaF} & \textbf{Kong} & \textbf{T5} \\
\midrule
Frame F1 & 0.8705 & \textbf{0.9132} & 0.7045 \\
Note Offset F1 & 0.7725 & \textbf{0.8751} & 0.8046 \\
Note Offset Velocity F1 & 0.7372 & \textbf{0.8602} & 0.7978 \\
Melody IOI & 0.2645 & \textbf{0.5300} & 0.2638 \\
Accompaniment IOI & 0.2232 & 0.3725 & \textbf{0.4377} \\
Melody KOR & 0.5202 & \textbf{0.6330} & 0.3450 \\
Bass KOR & 0.2979 & \textbf{0.6298} & 0.3182 \\
Ratio KOR & 0.5100 & \textbf{0.6153} & 0.3480 \\
Cloud Diameter & 0.7243 & \textbf{0.8301} & 0.7472 \\
Cloud Momentum & \textbf{0.2463} & 0.2152 & 0.1676 \\
Dynamics & 0.5501 & \textbf{0.6503} & 0.6355 \\
PEAMT \cite{ycart2020investigating} & \textbf{0.6304} & 0.6179 & 0.6116 \\
\bottomrule
\end{tabular}

    \captionsetup{width=.95\columnwidth}
    \caption{Model performance measured by standard metrics, our musically informed metrics, and PEAMT \cite{ycart2020investigating} on the MAESTRO subset of our evaluation set}
    \label{tab:maestro_all_metrics_res}
\end{table}

Generally, it can be observed that the Kong model performs the best across most metrics. This implies that most of our metrics, overall, correlate with the performance ranking as measured on the standard metrics. 
\rev{
Furthermore, it can be seen that the two Magenta models perform considerably different when measured against frame-level F1 score, yet this difference becomes less pronounced when evaluated on note-level metrics, which would suggest better performance of the T5 model. 
Comparing these results to the model performance as evaluated on our set of metrics, however, reveals that while both models perform similarly on onset time prediction, T5 is worse at adequately capturing note durations but better in estimating MIDI velocity and the overall loudness ratio between voices.
}
Lastly, we can observe that the perceptually informed PEAMT metric correlates most with the frame-level and harmony metric \emph{Cloud Momentum}, which might suggest (if PEAMT is indeed a veridical listening model) that listeners place relatively high importance on harmonic context.

\begin{figure}[ht]
  \includegraphics[alt={Relationship between model performance evaluated on note-offset-velocity F1 score and our proposed dynamics measure.},width=0.95\columnwidth]{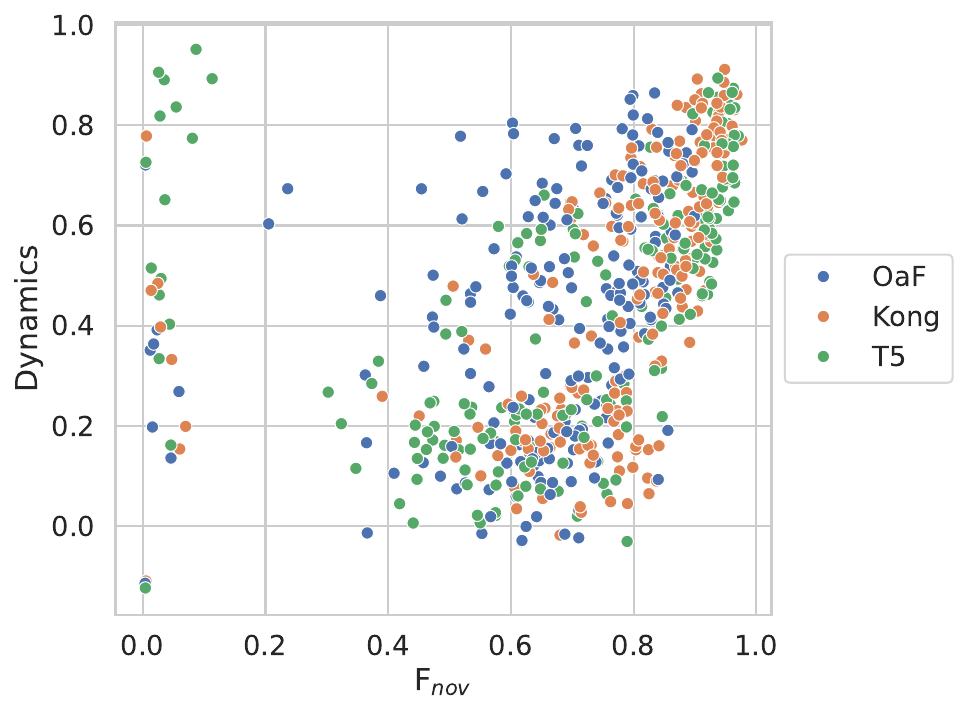}
  \captionsetup{width=.95\columnwidth}
  \caption{Relationship between model performance evaluated on note-offset-velocity F1 score and our proposed dynamics measure.}
  \label{fig:dynamics_vs-fnov}
\end{figure}

\begin{table*}[b!]
    \small
    \centering
    \def\arraystretch{1.44}
\small
\begin{tabular}{@{\extracolsep{4pt}}l l ccc ccc ccc@{}}
\toprule
 &  & \multicolumn{3}{c}{$\mathbf{frame}$} & \multicolumn{3}{c}{$\mathbf{note_{off}}$} & \multicolumn{3}{c}{$\mathbf{note_{off-vel}}$} \\
 \cmidrule(lr){3-5} \cmidrule(lr){6-8} \cmidrule(lr){9-11} \textbf{split} & \textbf{model/audio} & \textbf{OaF} & \textbf{Kong} & \textbf{T5} & \textbf{OaF} & \textbf{Kong} & \textbf{T5} & \textbf{OaF} & \textbf{Kong} & \textbf{T5} \\
 \cline{1-11}

\multirow[t]{2}{*}{train} & MAESTRO & 0.8803 & 0.9201 & 0.7259 & 0.7890 & 0.8910 & 0.8222 & 0.7530 & 0.8768 & 0.8167 \\
 & Disklavier & 0.8182 & 0.8503 & 0.6155 & 0.6630 & 0.7588 & 0.6371 & 0.6067 & 0.6836 & 0.5752 \\
\cline{1-11}
\multirow[t]{2}{*}{validation} & MAESTRO & 0.8401 & 0.8931 & 0.6544 & 0.7407 & 0.8631 & 0.8069 & 0.7077 & 0.8488 & 0.7992 \\
 & Disklavier & 0.7694 & 0.8675 & 0.6092 & 0.6261 & 0.8233 & 0.7335 & 0.5807 & 0.7568 & 0.6703 \\
\cline{1-11}
\multirow[t]{2}{*}{test} & MAESTRO & 0.8522 & 0.8995 & 0.6548 & 0.7306 & 0.8244 & 0.7401 & 0.6963 & 0.8068 & 0.7290 \\
 & Disklavier & 0.8046 & 0.8525 & 0.6046 & 0.6063 & 0.7235 & 0.6248 & 0.5550 & 0.6534 & 0.5567 \\
\cline{1-11}
\bottomrule
\end{tabular}
    \captionsetup{width=.95\textwidth}
    \caption{Frame-, note-offset, and note-offset-velocity F1 score results computed on our evaluation set, grouped per data set split, evaluated model and piano / audio environment.}
    \label{tab:eval_benchmark_results}
\end{table*}

We continue our discussion in Figure \ref{fig:composerwise_MAESTRO}, which compares the note-offset F1 score per composer (averaged over pieces and performers) and model to our musically informed timing, articulation and harmony metrics. We can see again that the Kong model performs best across all composers and metrics except for the \emph{Accompaniment IOI} timing and the \emph{Cloud Momentum} harmony metrics. The fact that it performs better on the \emph{Bass KOR} articulation metric, but poorly on those timing accompaniment and harmony metrics might suggest that this model detects many out-of-key extra notes that both erroneously influence the IOI sequence on the accompaniment part, and the estimation of the tonal context. \rev{Interestingly, we can can also observe, as a general trend, that the F1 score somewhat deteriorates with increasing virtuoso and challenging musical repertoire. This general deterioration with increasing musical difficulty is not reflected correspondingly by our metrics, which show more variation with respect to different composers and aspects of the underlying music: While the F1 score for Bach, Mozart, Haydn and Schubert all suggest a near-perfect transcription, our metrics indicate more diverse results, and e.g., suggest poor(er) accuracy (and therefore reliability in a performance study context) in timing aspects.} Another illustrative example can be found in the case of Chopin: here again the F1 score (particularly of the Kong model) would suggest a highly accurate transcription output, while our metrics reveal that the expressive dimension of articulation is not well captured. Lastly, we can observe again that the PEAMT and harmony metric \emph{Cloud Momentum} show a similar trend for most composers, suggesting a greater weight of the harmonic context in that trained metric.

We conclude our discussion by examining the dynamics aspect. Figure \ref{fig:dynamics_vs-fnov} illustrates the relationship between the note-offset-velocity F1 score and our proposed \emph{Dynamics} metric. While both metrics show a weak correlation (Pearson $r=0.21$), the figure also indicates that our metric evaluates dynamics in a more differentiated way and leads to a wider range of evaluation results than the standard metric. Note that our metric only evaluates the dynamics aspect, in particular how well the overall balance in loudness between different voice streams is preserved in a transcription. It does not account for onset, offset and pitch information, which also explains the results in the very left part of the figure that score low on F1 score but high on our metric.

\section{Out-of-distribution inference}\label{sec:ood_inference}

In this section we illustrate the problem of out-of-distribution performance of the models analysed. We believe that this is an important aspect to emphasize, as transcription models are ultimately intended to be (and have been) used on real-world audio performances \cite{kong2020giantmidi, zhang2023atepp}. 

We approach the problem in two stages: First, we elucidate the problem by performing a short evaluation of the three analysed models on real-world recordings using only the standard IR metrics. Second, we simulate more challenging real-world environments using different levels of noise and reverberation, and evaluate the analysed models again using the standard and our proposed metrics, where we highlight how our musically informed metrics can reveal aspects that the standard metrics would otherwise have missed.

\subsection{Generalization on real-world recordings}

Table \ref{tab:eval_benchmark_results} shows the mean frame- and note-level F1 scores per model and per piano/acoustic recording environment on the three splits of the MAESTRO dataset (as they are officially defined).\footnote{We note that for the two Magenta models, OaF and T5, our evaluation results do not come close to the reported ones in \cite{hawthorne2019enabling} and \cite{hawthorne2021sequence}. The differences are particularly pronounced in the note-level F1 scores.}




We group the results per split to test whether the analysed models would perform worse on out-of-distribution recordings of performances (pieces) from the test set compared to those from the train/validation sets.
Next we conduct a Kruskal Wallis ANOVA\cite{vargha1998kruskal} to test for differences between frame-, note-offset and note-offset-velocity F1 scores, grouping the evaluation scores each by \textit{split} and by \textit{audio environment} and comparing each model separately. For each ANOVA we use a significance threshold of $\alpha=0.05$. The ANOVA on the \textit{audio environment} dimension show a statistically significant difference ($p<0.05$) between the MAESTRO and Disklavier audio recordings for all three analysed models \rev{and all F1 metric levels considered}.

The ANOVA on the \textit{split} dimension yields more differentiated results:  For all models, the frame-level F1 scores are significantly different, and there are no statistically significant differences in the note-offset-velocity F1 scores. \rev{For the Kong and OaF models, the note-offset-score is significantly different depending on the split, whereas the T5 shows no significant differences. These results suggest that the most musically meaningful metric from the current set of standard metrics\cite{hawthorne2017onsets} might not sufficiently capture overfitting tendencies.}




\subsection{Evaluation on perturbed audio recordings}

Similar as in Section \ref{sec:demonstration}, we again compare our musically informed metrics to the standard IR and the PEAMT metrics, however, this time on a set of more challenging audio recordings. To this end, we choose six (MAESTRO and \textit{Disklavier}) audio recordings which we artificially perturb by introducing reverberation and synthetic noise. 
We use three Impulse-Response filters, modelling short, medium and long reverberation times (RT60@1kHz $\in \{0.19, 1.85, 10.5\}$ seconds) and sourced from the OpenAIR\footnote{\url{https://www.openair.hosted.york.ac.uk}} database.
We further add white noise into the recordings at three different Signal-to-Noise Ratio levels ($\text{SNR}_{dB} \in \{24, 12, 6\}$).
Following a factorial design with these two independent variables, each with four levels, we first perturb the audio recordings on all conditions, and transcribe these recordings using all three analysed models. Following this procedure, we obtain 284 transcribed MIDI performances.\footnote{Note that 6 pieces x 4 noise levels x 4 reverberation levels x 3 models yield 288 transcriptions, but 4 recordings (each at the two higher most of either reverberation and/or noise levels) 
resulted in empty transcriptions (zero predicted note events) by the T5 model, and are therefore excluded from the evaluation.}

\begin{figure}[t]
  \includegraphics[alt={Performance degradation analysis as evaluated on the note-offset F1 score and our proposed \emph{Melody IOI} timing and \emph{Cloud Momentum} harmony metrics.},width=0.95\columnwidth]{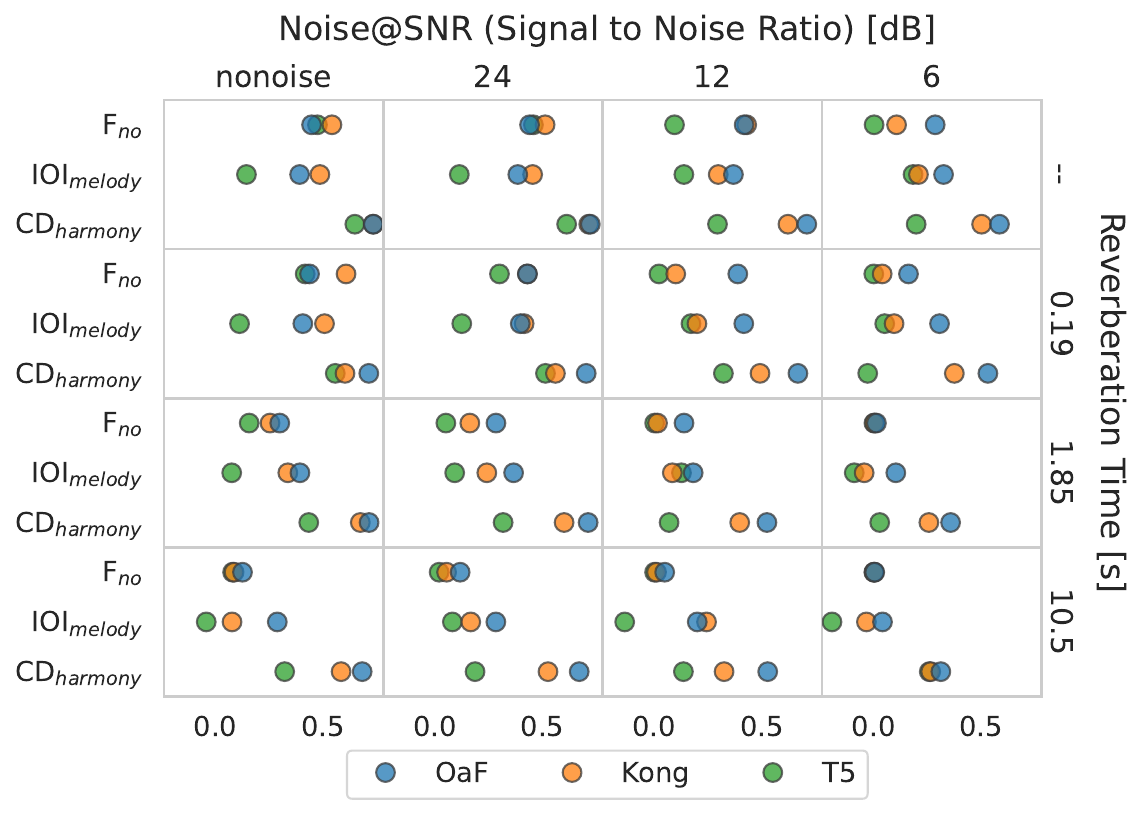}
  \captionsetup{width=1\columnwidth}
  \caption{Performance degradation measured by note-offset F1, \emph{Melody IOI} and \emph{Cloud Momentum} metrics.}
  \label{fig:exp2_revnoise_fig}
\end{figure}

Each grid cell in Figure \ref{fig:exp2_revnoise_fig} compares the mean note-offset F1 scores per model to the \emph{Melody IOI} timing metric and \emph{Cloud Momentum} harmony metric, where grid rows represent increasing reverb levels, and grid columns represent increasing noise levels.
Generally, it can be observed that the performance range of models as measured by the F1 score is notably reduced compared to our metrics, indicating that our metrics possess higher discriminative capacity than the standard ones.

As expected, the inference performance of all three analysed models deteriorates with increasing noise and reverberation levels, though the deterioration is less pronounced on the noise than on the reverberation axis. 
Furthermore, analysing the results on the timing metric \emph{Melody IOI} suggests that the model by Kong predicts onset times worse with increasing noise levels, while the onset times prediction by OaF seems to be more resistant to this form of perturbation. Finally, the results measured on the harmony metric \emph{Cloud Momentum} suggest that the overall harmonic context is relatively well preserved at higher perturbation levels by the OaF and Kong models, and less so by the T5 model.

\section{Conclusion}\label{sec:conclusion}

In this study, we investigated two aspects that are commonly neglected in the evaluation of transcription models: (i) limited explanatory power of the standard IR evaluation metrics with respect to the underlying musical material, and (ii) poor inference on out-of-distribution data. We study both problems in the context of solo piano transcription, and, in addressing the first aspect, propose a set of musically informed metrics designed to capture more musically relevant information, particularly for the context of computational studies of expressive performance.

We demonstrated our metrics on transcriptions obtained by three state-of-the-art piano transcription models on a subset of the MAESTRO dataset, the de-facto standard train and test set for current transcription models, and highlighted musical dimensions for which they provide more informative value than the standard information retrieval metrics. We have further illustrated the lack of generalization with respect to the acoustic environment, both on real-world and perturbed audio recordings.

\ch{


Future work in this direction may include an extension and further validation of our new musically informed metrics, in order to capture additional qualities of expressive performance, potentially by making use of score alignment information.
Additionally, a listening study with human experts could help further investigate the perceptual validity of our proposed metrics.
}

\section{Acknowledgments} 
This work receives funding from the European Research Council (ERC), under the European Union's Horizon 2020 research and innovation programme, grant agreement No.~101019375 (\textit{Whither Music?}). The LIT AI Lab is supported by the Federal State of Upper Austria.


\bibliography{ISMIRtemplate}

\end{document}